\documentclass[12pt,preprint]{aastex}
\usepackage{graphicx}

\shorttitle{A Theoretical Investigation on the GRB Host Galaxies}
\shortauthors{J. Mao}

\begin{document}


\title{A Theoretical Investigation on the Gamma-ray Burst Host Galaxies}


\author{Jirong Mao}
\affil{INAF-Osservatorio Astronomico di Brera, Via Bianchi 46, I-23807, Merate (LC), Italy}
\affil{Yunnan Observatory, Chinese Academy of Sciences, Kunming, Yunnan Province, 650011, China}
\affil{Key Laboratory for the Structure and Evolution of Celestial Objects, Chinese Academy of Sciences}

\email{jirong.mao@brera.inaf.it}

\begin{abstract}
Long-duration gamma-ray bursts(LGRBs) are 
believed to be linked with the star formation. We adopt a 
galactic evolution model, 
in which the star formation process inside the virialized 
dark halo at given redshift can be achieved. In this paper, 
the gamma-ray burst(GRB) host galaxies are assumed to be the star-forming 
galaxies within the small dark halos. The star formation 
rates(SFRs) in the host galaxies of LGRBs at different 
redshifts have been derived from our model with the 
galactic evolutionary time about a few times of $10^7$ yr 
and the dark halo mass of about $5\times 10^{11}M_\odot$. The 
related stellar masses, luminosities and metallicities of these 
hosts are estimated as well. We further calculate the X-ray 
and optical absorption of GRB afterglow emission. From our 
model calculation, at higher redshift, the SFR of host galaxy 
is larger, the absorption in X-ray band and optical band of GRB 
afterglow is stronger, in the condition that the dust and metal 
components are released locally, surrounding the GRB environment. 
These model predictions are compared with the {\it Swift} and 
other observational data. At lower redshift $z<1$, as the 
merger and interaction of some host galaxies are involved, one 
monolithic physical process is not sufficient to fully explain 
all kinds of observed phenomena.
\end{abstract}


\section{Introduction}
Gamma-ray burst(GRB), the most violent explosion cosmic source, has been identified as the cosmological event since 1997(van Paradijs et al. 1997; Metzger et al. 1997).
Recently GRB 090423 has been explored at high redshift above 8(Salvaterra et al. 2009a; Tanvir et al. 2009). 
The long-duration GRB(LGRB) progenitors are proposed to be the massive collapsing stars (Woosley 1993; Kumar, Narayan \& Johnson 2008).
Some long bursts have been observed to be associated with supernova events(Hjorth et al. 2003; Stanek et al. 2003; Malesani et al. 2004; Mazzali et al. 2006; Xu et al. 2008), 
hence having a common star-forming origin(Paczynski 1998). Indeed, long GRBs can be found in the star formation galaxies and these galaxies are dominated 
by the young stellar population(Christensen et al. 2004). In general, GRBs favor a metal-poor environment(Fynbo et al. 2006; Kewley et al. 2007) and the 
hosts have low stellar masses(Wiersema et al. 2007). Jakobsson et al. (2005) proposed that GRB host galaxies, at least those high redshift($z>2$) hosts, trace the star 
formation of the universe in an unbiased way. The high global star formation rate(SFR) history at redshift larger than 6 (Hopkins \& Beacom 2006; Yan et al. 2009) indicates 
the possibility of high-redshift GRB production and the detection of host galaxies. From the research of Y\"{u}ksel et al. (2008) and  Kistler et al. (2009), there 
could be a link from star formation to the GRB production in the high redshift universe, in which the GRB luminosity function is involved. Moreover, the evolution of the GRB 
luminosity function has been investigated by Salvaterra et al. (2009b). All of these evidence provide the strong clue to study the intrinsic link from SFR to GRB production 
and the possible evolutionary properties of GRBs and their hosts.  

The grains and metals produced by the host galaxy will take effects on the GRB afterglow emissions. Thus, the GRB progenitors and their environments can be expressed by the  
absorption features of GRB afterglows. The heavy attenuation in the X-ray band has been given in the statistic results from Campana et al. (2010), indicating a dense 
surrounding environment of those GRBs. In the mean while, it is also interesting to understand whether this kind of strong attenuation is intrinsically evolved with redshift. 
On the other hand, the characteristics of the corresponding absorption in the optical band are still under debate. Although the approximate dust 
extinction law of GRB host galaxies has been given by Chen, Li \& Wei (2006) and Li et al. (2008), in order to have an explanation of dust obscuration and especially to 
interpret some X-ray detected but optical faint bursts(so-called dark bursts, Akerlof \& Swan 2007; Kann et al. 2007; Perley et al. 2009), the physical origin associated 
with the star formation and galactic evolution should be studied in an unified scenario.

In this paper, we specify one physical model of star-forming and metal-poor galaxies being as the hosts of long GRBs, exploiting the physical recipes from Granato et al. 
(2004). In the general scenario of Granato et al. (2004), at each redshift bin, the SFR and galaxy mass in the given dark halo potential well have been 
calculated, with the effects on the kinetic feedback of supernova and central black hole. Under this framework, the different evolutionary stages of galaxies and the central 
black holes with different physical conditions have been investigated(e.g., Cirasuolo et al. 2005 about the properties of E/S0 galaxies; Lapi et al. 2006 about the active
galactic nucleus luminosity function; Granato et al. 2006 about the submillimeter galaxies). In particular, Mao et al. (2007) calculated the UV luminosities and the relative 
dust attenuation 
in the star-forming and metal-poor galaxies, Lapi et al (2008) estimated the long GRB progenitor rates and redshift distribution. Since the updated X-ray/optical 
observations on the GRB afterglows and host galaxies have been performed sequentially by Castro Cer\'{o}n et al. (2008), Evans et al. (2009), 
Savaglio, Glazebrook \& Le Borgne (2009), Levesque et al. (2009a) and Fynbo et al. (2009), in this context, it is necessary to further compare some properties calculated by 
our model with these updated observational data. We extend the former calculation from Mao et al. (2007), attempting to understand the physical origin of the long GRB 
production and the GRB environment, especially, we reveal that some properties from afterglow emissions and GRB hosts have shown the possible intrinsic cosmological evolution.

Throughout the paper, we adopt cosmological parameters: $h=0.7$, $\Omega_M=0.3$, and $\Omega_\Lambda =0.7$. 

\section{Model Predictions}
\subsection{Model Review}
In the following we report briefly on some key aspects under the framework of star formation in the protogalaxies from our recipes(see also Appendix A of Mao et al. 2007 and 
Lapi et al. 2008 for details). In general, the star formation process and the central black hole growth are inside the given virialized dark halo with the mass $M_{halo}$. 
The cooling gas will infall toward the center of the dark halo and form into the stars and galaxy; in the mean while it will be heated by the central black hole activity. 
Thus, the total infalling gas $\dot M_{inf}=-\dot M_{cond}-\dot M^{BH}_{inf}$ includes two parts: one is the condensation gas toward the center of the 
dark halo $M_{cond}$, the other is the gas removed by the central black hole activity $M^{BH}_{inf}$. The condensation timescale $t_{cond}$ is the maximum between the dynamic 
timescale and the cooling timescale at the halo virial radius. Thus, the cold gas $\dot M_{cold}=\dot M_{cond}-(1-R)\dot M_{\ast}-\dot M_{cold}^{SN}-\dot M_{cold}^{BH}$, where
$\dot M_{\ast}=M_{cold}/t_{\ast}$ is the SFR and R is the fraction of gas transferred to the cold component by the evolved stars. Adopting the initial mass 
function(IMF) by Romano et al. (2002), we have $R\sim 0.3$, $\dot M_{cold}^{SN}$ and $\dot M_{cold}^{BH}$ are the feedback from supernova and central black hole respectively. 
Therefore, with the scaling approximation, we have SFR $\dot M_{\ast}(t)=M_{inf}(0)(e^{-t/t_{cond}}-e^{-s\gamma t/t_{cond}})/t_{cond}(\gamma-1/s)$, where 
t is the evolutionary time, $\gamma=1-R+\beta_{SN}$, $\beta_{SN}$ is the ratio between star formation feedback by supernova and SFR, 
$s\sim t_{cond}/t_{\ast}\sim 5$. In a virialized dark matter halo, the total gas $M_{inf}(0)$ is about 18\% of the dark halo mass. 
The condensation timescale can be estimated as $t_{cond}=4\times10^8((1+z)/7)^{-1.5}(M_{halo}/10^{12}M_\odot)^{0.2}~yr$. The central black hole quenches the star formation in 
the halo effectively after the time about $t_{BH}=2.5\times 10^8((1+z)/7)^{-1.5}F(M_{halo}/10^{12}M_\odot)~yr$, where $F(x)=1$ for $x\ge 1$ and $F(x)=x^{-1}$ for $x\le 1$. 
In other words, the cooling gas inside the virialized dark halo forms into stars and galaxy, the star formation process begins and persists with a relatively high rate until a
few times of $10^8$ yr so that the central seed black hole growth is enough to be a supermassive black hole and it shines as a quasar. After that, the central black hole 
will release the kinetic feedback, heat the cold gas, and quench the star formation. 

In this paper, the GRB host is believed to be the young, star-forming galaxies. During the forming time about a few times of $10^7$ yr, the star formation process is 
violently on-going. The masses of these host galaxies are in general less than $10^{10}M_{\odot}$(Savaglio et al. 2009). About these GRB host galaxies,
three physical inputs emphasized below can decide the whole recipes: (1) redshift z: at different redshifts, the star formation and galactic evolution processes are 
different; (2) dark halo mass $M_{halo}$: as Mao et al. (2007) and Lapi et al. (2008) pointed out, the host dark halos in which GRBs occur are relatively small, usually, they 
are less than $10^{12} M_{\odot}$. In this paper, we select $5\times 10^{11} M_{\odot}$ as a reference value; this is consistent with the simulation results of 
Courty et al. (2007) and Campisi et al. (2009); 
(3) evolutionary time t: the GRB host galaxy is in the initial stage of the galaxy evolution; this initial time t is about a few times of $10^7$ yr, less than about a few 
times of $10^8$ yr. This value is supported by the observations from Th\"{o}ne et al. (2008) and 
Han et al. (2010). It is noted that at this stage the central black hole seed does not have enough growth to be a supermassive black hole; thus, the quasar feedback can be 
ignored in our calculations. The evolutionary time t can be roughly estimated as the starburst time in one starburst galaxy as well and the central black hole activity takes 
negligible effects on the galactic evolution. Thus, after these three inputs are given, the SFR in the GRB host galaxies can be decided.

\subsection{Results}
We begin the procedure from the SFR and galactic mass calculation of the hosts, the B-band absolute magnitudes of the hosts can be derived from the empirical relation of 
Savaglio et al. (2009). We can also obtain the metallicity distribution by applying the mass-metallicity relation of Savaglio et al. (2005). 
Through the transition from UV band attenuation $A_{uv}$ to the dust absorption $A_v$, we follow the recipes of Mao et al. (2007), in which the results of 
Calzetti et al. (2000) have been adopted, as it is suitable for the high redshift star-forming galaxies. We further transfer $A_v$ to X-ray column density $N_{H,x}$ using the 
average value obtained by Schady et al. (2010). With the SFR and metallicity properties, the $A_v$ distribution can be derived as well.

\subsubsection{Star Formation and Metallicity of GRB Host Galaxies}
\begin{figure}
\includegraphics[height=16cm,angle=-90]{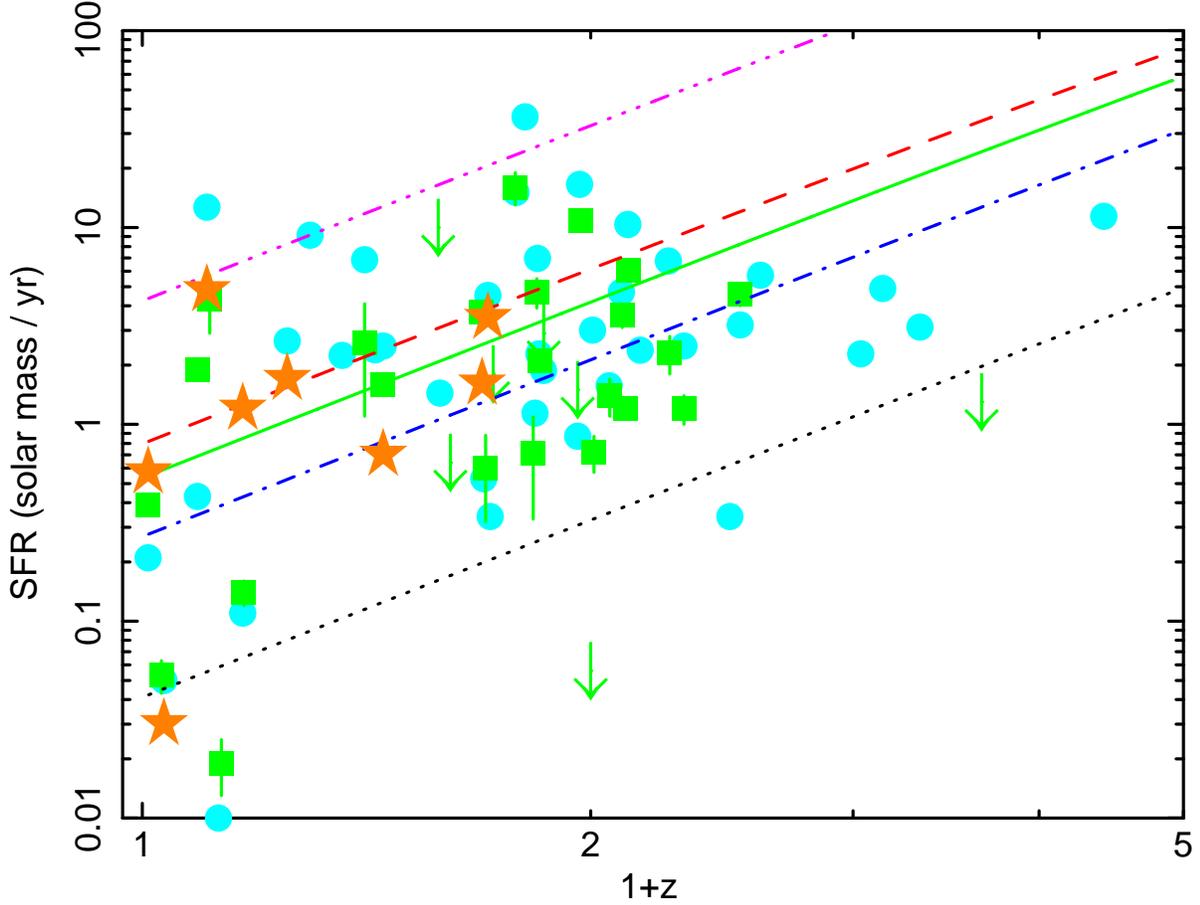}
\caption{SFRs of GRB host galaxies. The dash-dotted line, the solid line and the dashed line denote the prediction results by the model under the condition of dark halo 
mass $5.0\times 10^{11}M_\odot$, their corresponding galactic evolutionary times are $2.5\times 10^7~yr$, $5.0\times 10^7~yr$ and $7.5\times 10^7~yr$ respectively. The dotted 
line
and the dash-double-dotted line are the lower and upper limits given from the model, with the two set parameters (halo mass $1.0\times 10^{11}M_\odot$, galactic timescale 
$1.0\times 10^7~yr$ and halo mass $5.0\times 10^{12}M_\odot$, galactic timescale $1.0\times 10^8~yr$ respectively). The observational SFR values taken 
from Savaglio et al. (2009) are shown as dots, from Castro Cer\'{o}n et al. (2008) as squares, and from Levesque et al. (2009a) as stars.\label{f1}}
\end{figure}

As we assume that long GRBs occur inside the young and star-forming galaxies, the star formation process plays a key role on the environment of GRB production. With the 
reference values about $5.0\times 10^{11}M_\odot$ for a dark halo mass and $5.0\times 10^7~yr$ for the galactic evolution time, we have reproduced the SFR at each redshift, 
the stellar mass of host galaxy can be derived as SFR multiplied by the galactic formation timescale. The galactic formation timescale can be defined as 
$t_g=t\cdot ((1+z)/7)^{-1.5}(M_{halo}/10^{12}M_\odot)^{0.2}~yr$, which has the same index numbers of condensation timescale. We have shown the SFR of GRB host galaxies at each
redshift in Fig. \ref{f1}. In this figure, the observational SFR values that spread over a wide range, from 0.01$M_\odot/yr$ to about $10M_\odot/yr$ within the redshift
bin $0<z<1$, are shown. Our model predicts in general that SFR values are larger toward the redshift larger than 1. 

\begin{figure}
\begin{center}
\includegraphics[height=16cm,angle=-90]{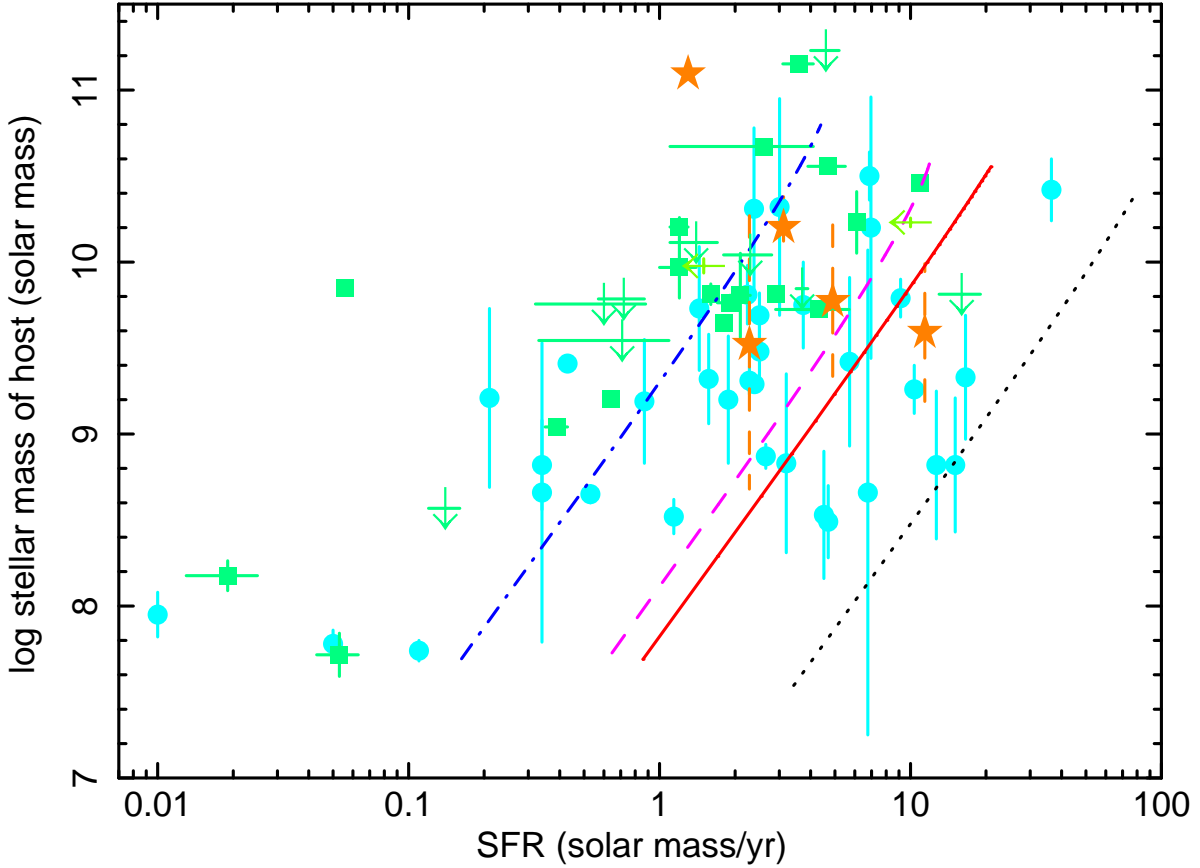}
%lcIab.dat
\caption{Stellar mass as a function of SFR. The model prediction range is dependent on the adopted galactic formation timescale. The model predicted solid 
and dash lines are in the condition of redshift 1.0, with the halo masses of $5\times 10^{11}M_\odot$ and $1\times 10^{11}M_\odot$, respectively. The galactic timescale 
ranges are $t_g\sim 10^{7.0-8.5}~yr$ for the results denoted by the solid line, and $t_g\sim 10^{7.7-9.0} yr$ for the results denoted by the dashed line. The prediction at 
redshift 0.1 is denoted as the dash-dotted line, the result is under the condition of halo mass $1.0\times 10^{11}M_\odot$ and the galactic timescale range 
$t_g\sim 10^{7.5-9.2}~yr$.
At redshift 3.0, we get the results denoted by the dotted line, under the condition of $5.0\times 10^{11}M_\odot$ and the timescale range $t_g\sim 10^{6.7-8.2}~yr$. 
The near-infrared-selected GRB hosts(denoted as squares; Castro Cer\'{o}n et al. 2008) have larger mass values than the hosts selected by the optical observations(denoted as 
dots; Savaglio et al. 2009).  
We specify the data at redshift larger than 2, shown as stars in the plot. 
\label{f2}}
\end{center}
\end{figure}

The relation between the SFR and the stellar mass of GRB host galaxy is given in Fig. \ref{f2}. This possible correlation is also mentioned by Savaglio et 
al. (2009). In our model, we illustrate that this correlation at given redshift with a given dark halo mass could be due to the growth of the host protogalaxies under the 
certain galactic formation timescale. However, there is no straightforward relation shown by the observational data in Fig. \ref{f2}. We note that within the different dark 
halos the SFRs and the stellar masses are different. Furthermore, some galaxies with relatively larger masses at lower redshift might have experienced twice or
even more times of starbursts during their lifetimes. Especially, it is found easily in the plot that the infrared-selected host galaxies have larger stellar masses.   
This complicated situation indicates that at low redshift, the GRB host galaxies might not have a monolithic evolutionary process. The starburst triggered by merging 
or interaction can happen as well. The further discussion about the host galaxies in the low-redshift universe will be given in Section 3.   

\begin{figure}
\begin{center}
\includegraphics[height=16cm,angle=-90]{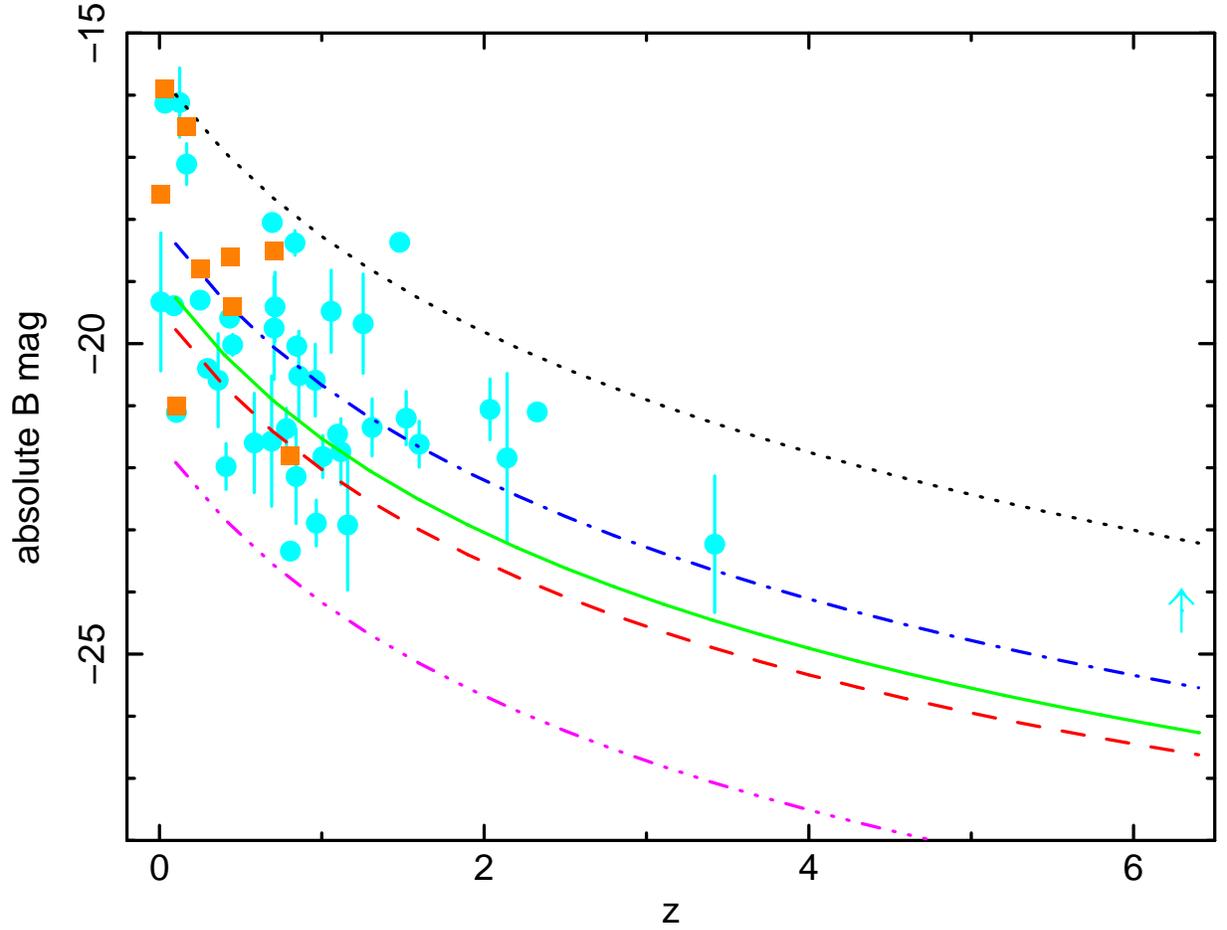}
%lcIab.dat
\caption{B band absolute magnitudes of GRB host galaxies. All the model predictions are the same as those in Fig. 1. The observational data from
Savaglio et al. (2009) are marked by dots while those from Levesque et al. (2009a) are marked by squares.\label{f3}}
\end{center}
\end{figure}

In the work of Courty et al. (2007), the ratio between SFR and B-band luminosity of GRB host galaxy has been investigated with the observational data from Christensen 
et al. (2004). Here, we adopt the observational results by Savaglio et al. (2009). Assuming the correlation between SFR and B-band absolute magnitude, we find that the
data are well described by the scaling relation $logSFR=-(0.36\pm 0.01)M_B-(6.72\pm 0.27)$. Therefore, using this scaling relation we have the B-band absolute magnitude of GRB
host galaxies in each redshift bin, the results are shown in Fig. \ref{f3}. It is suggested by Malesani et al. (2009) that at higher redshift the GRB host galaxies could be 
brighter. In our model, we point out that this is the result coming from the intrinsic SFR redshift distribution.  

\begin{figure}
\begin{center}
\includegraphics[height=16cm,angle=-90]{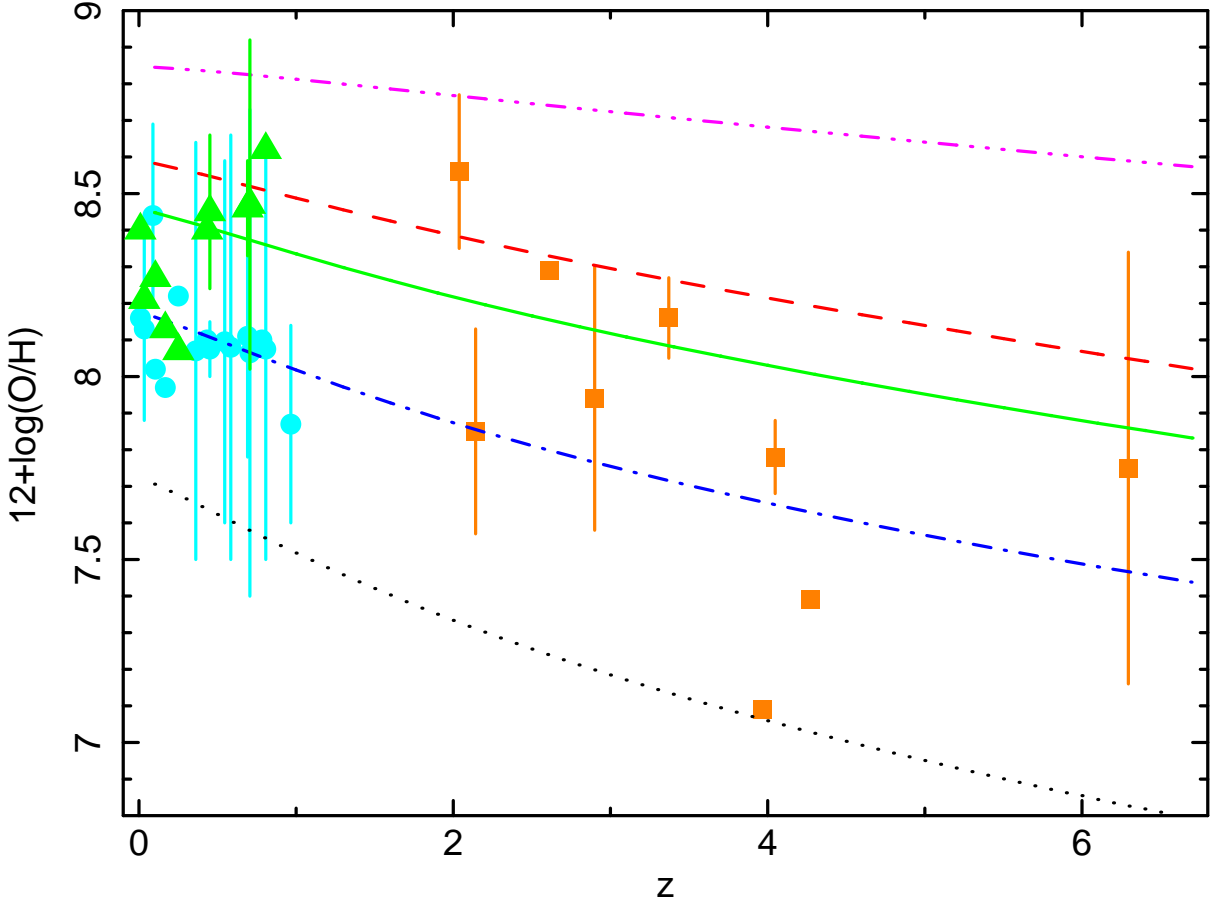}
\caption{Metallicity distribution with the normalization $12+log(O/H)=log(Z/Z_\odot)+8.69$. The model predictions are the same as those in Fig. 1. The average values 
from Savaglio et al. (2009) are denoted by dots, and the data from Levesque et al. (2009a) are marked by triangles. The metallicity values of GRB-damped Ly$\alpha$ systems
(Savaglio 2006) are presented by squares.\label{f4}}
\end{center}
\end{figure}

In order to investigate the possible metallicity distribution, in this paper, assuming that the mass-metallicity relation and its 
redshift evolution(Savaglio et al. 2005) are valid for the GRB host galaxies as well, after calculation on the stellar masses of GRB hosts, we obtain the metallicity evolution
as shown in Fig. \ref{f4}. The metallicity values of the host galaxies slightly decrease toward the higher redshift. This finding is consistent with that obtained by 
Li (2008). 
We caution that the metal-poor case is not a necessary input condition in our model; thus, the metallicity may not be the essence for GRB production. Further metallicity 
estimations of GRB host galaxies are given in Section 3.  

\subsubsection{Afterglow Absorptions}
\begin{figure}
\begin{center}
\includegraphics[height=14cm,angle=-90]{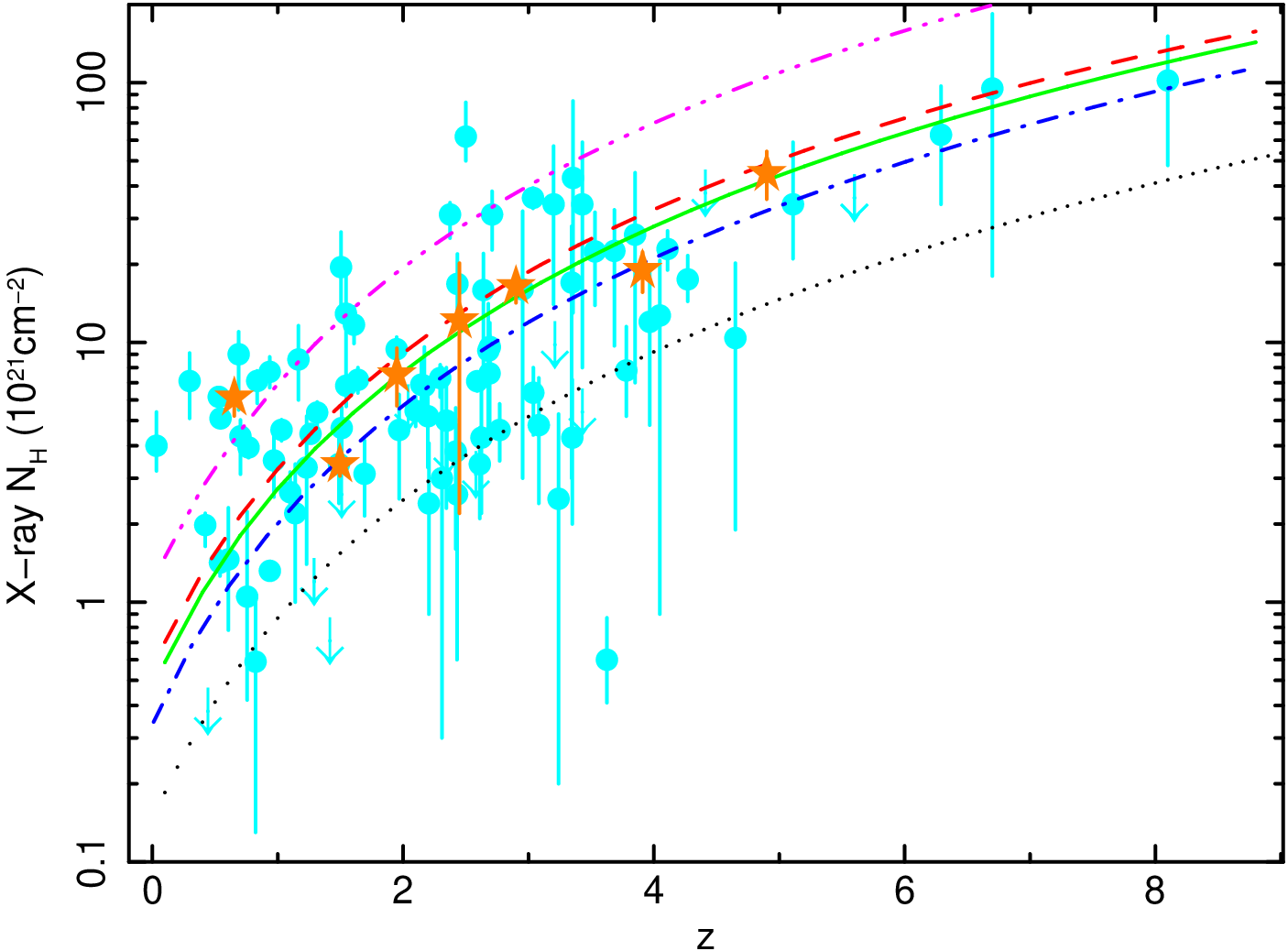}
\includegraphics[height=14cm,angle=-90]{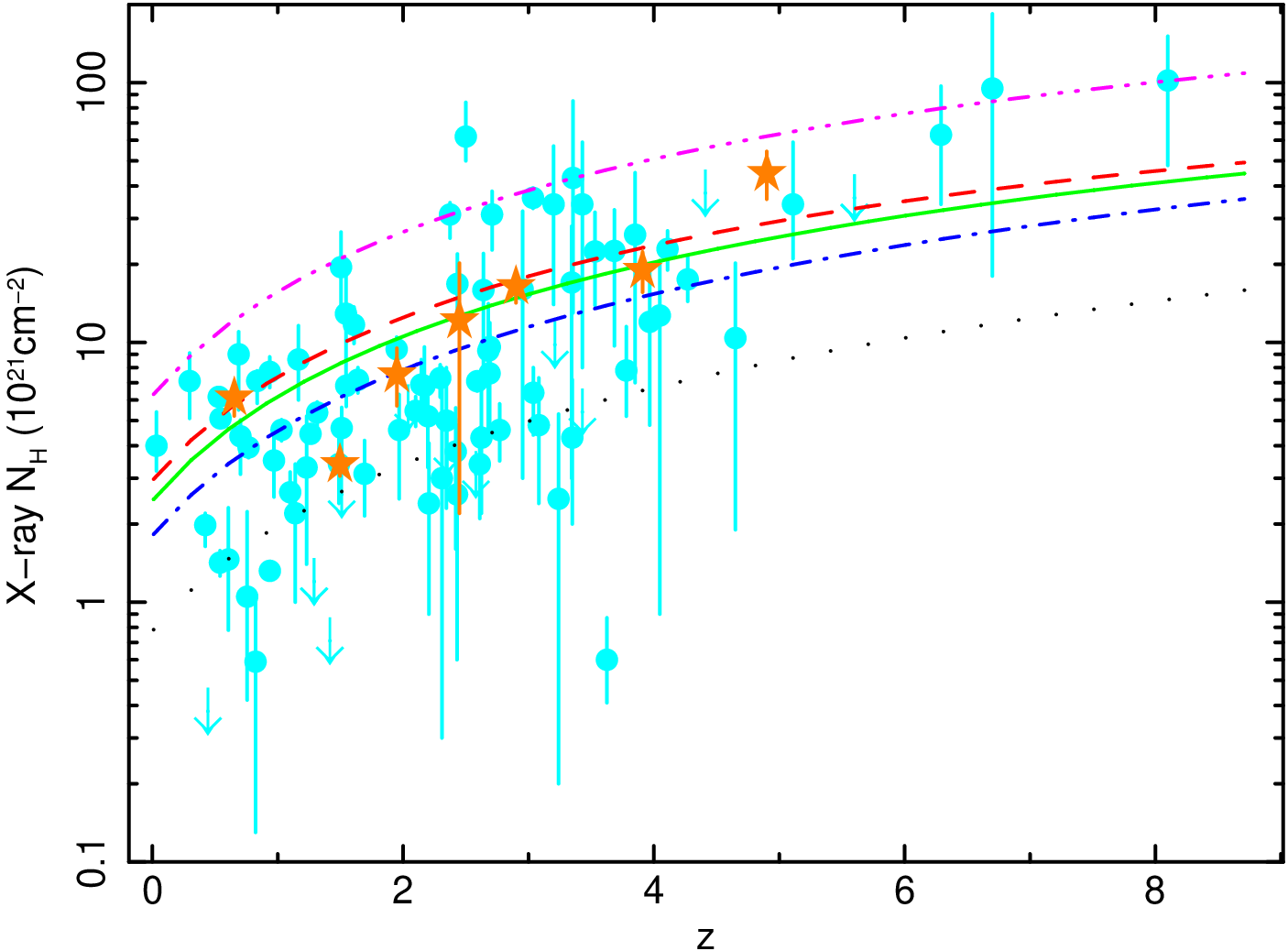}
\caption{X-ray absorption $N_{H,x}$ distribution with redshift. All the model predictions are the same as those in Fig. 1. The observational data are
taken from Evans et al. (2009) and Campana et al. (2010). In the plot, the dark bursts identified by Perley et al. (2009) and Zheng et al. (2009) are labeled 
as stars.\label{f5} The results from the model shown in the upper panel (a) is the $N_{H,x}$ distribution with the possible selection effects. The model results with the 
depression of selection effects are shown in the lower panel (b).}
\end{center}
\end{figure}

X-ray Telescope(XRT), one of the instruments onboard the {\it Swift} satellite, has supplied the important X-ray data in the 0.3-10 KeV band for the GRB research. The 
Swift-XRT analysis has been performed automatically, and the spectral results for Swift-observed GRBs have been presented by Evans et al. (2009). Usually the X-ray 
spectrum is fitted by an absorbed power law. Thus, the X-ray photon index and the corresponding neutral hydrogen column density $N_{H,x}$ of each GRB can be achieved. We 
select each $N_{H,x}$ value of redshift-measured GRB and plot these values in Fig. \ref{f5}. 
According to the model described by Mao et al. (2007), the UV band absorption $A_{UV}=0.35(\dot M_\ast/M_\odot~yr^{-1})^{0.45}(Z/Z_\odot)^{0.8}$ is a function of  
SFR and metallicity. Following the calculation of Mao et al. (2007), we calculate from UV attenuation to $E(B-V)$ using $E(B-V)=A_{UV}/11$ by Calzetti et al. (2000). 
The results are in agreement with the observations(see Fig. 2 of Mao et al. 2007).
With $R_v=3.1$,  we obtained 
the dust attenuation $A_v$. With the data observed by Swift-XRT and Swift-UV/Optical Telescope(UVOT), Schady et al. (2007 \& 2010) modeled the spectral energy distributions 
and derived the ratio between $N_{H,x}$ and $A_v$. As the ratios derived from Schady et al. (2010) might be varied with the redshift, assuming the linear relation between 
redshift and the ratio with a logarithmic scale, we perform the linear regression on the data and obtain the optimized relation as 
$log(N_{H,x}/A_v~10^{21}cm^{-2})=1.24log(1+z)+0.79$ with the average standard deviation 0.37. 
Then we use this relation to transfer $A_v$ to $N_{H,x}$ and compare the results to the X-ray absorption data\footnote{As the XRT spectra are processed by the standard 
software XSPEC, in which the metallicity is fixed as the solar value, thus the observational data and the calculations of $N_{H,x}$ are all uniformed by the solar 
metallicity.} in Fig. \ref{f5} panel (a).

However, we note that the selection effects are included in the results of Evans et al.(2009) and Campana et al.(2010). As pointed out by Campana et al. (2010), at high 
redshift larger than 4, the intrinsic X-ray emissions suffer lower absorption; thus, the X-ray afterglows with low X-ray column densities are hard to be identified by 
Swift-XRT. On the other hand, as we use the 
data from Schady et al. (2010), although it was claimed that generally the selection effects on the distribution of host column densities are not significant, in our paper, 
in order to investigate the possible selection effects on our results, first, we check the possible $N_{H,x}-z$ relation and the $A_v-z$ relation respectively from the data of
Schady et al. (2010). We see that the correlation between $N_{H,x}$ and redshift has the efficient $r=0.67$ with null hypotheses 0.0006, this relation could be due 
to the selection effect mentioned above, but we do not find any possible relation between $A_v$ and redshift. As $N_{H,x}$ increases with redshift, $N_{H,x}/A_v$ also 
increases with redshift. Second, aiming to avoid this selection effect, we use the average value $<N_{H,x}/A_v>=3.3\times 10^{22} cm^{-2}$ given by Schady et al. (2010) to 
transfer $A_v$ to $N_{H,x}$ again. We plot the results in Fig. \ref{f5} panel (b). By using the mean value of $N_{H,x}/A_v$, the selection effect can be effectively depressed.

After the depression of selection effect, our model results still show a slight trend of X-ray absorption evolution. This evolution trend may be intrinsic. From our model, 
we see that the X-ray absorption is originally from the SFR. SFR has the redshift evolution as $SFR\sim (1+z)^{2.71}$. Under the assumption of solar metallicity,
we have the intrinsic X-ray attenuation $N_{H,x}\sim (1+z)^{1.22}$. Therefore, we conclude that the SFR redshift evolution is the dominant reason for the X-ray attenuation 
evolution shown in Fig. \ref{f5} (b). If we use the linear relation between $N_{H,x}/A_v$ and redshift, meaning the possible selection effects are included, we have the final 
results shown in Fig \ref{f5} (a).  We see that the intrinsic evolution plus the selection effects can fit the observational data of Evans et al. (2009) and Campana et al. 
(2010) well.  

Through the analysis above we clearly see, that the final results of GRB X-ray absorption are the calculations of intrinsic SFR redshift evolution, modified by the variation 
between $N_{H,x}/A_v$ and redshift. The later could be due to the selection effect. From Fig. \ref{f5}, we see that the observational data have large scatter. On the other
hand, our model provides the different values under the different dark halo masses and evolutionary time. Therefore, we also conclude that the large 
absorption is due to the longer galactic evolution time within the massive dark halo, while the small attenuation is due to the shorter galactic evolution time 
within the smaller dark halo. 
From the theoretical point of view, we confirm that the absorption is from the local environment of GRB, as suggested by Campana et al. (2010), Nardini et al. (2009), and 
Zheng et al. (2009) from the data analysis. 

\begin{figure}
\begin{center}
\includegraphics[height=16cm,angle=-90]{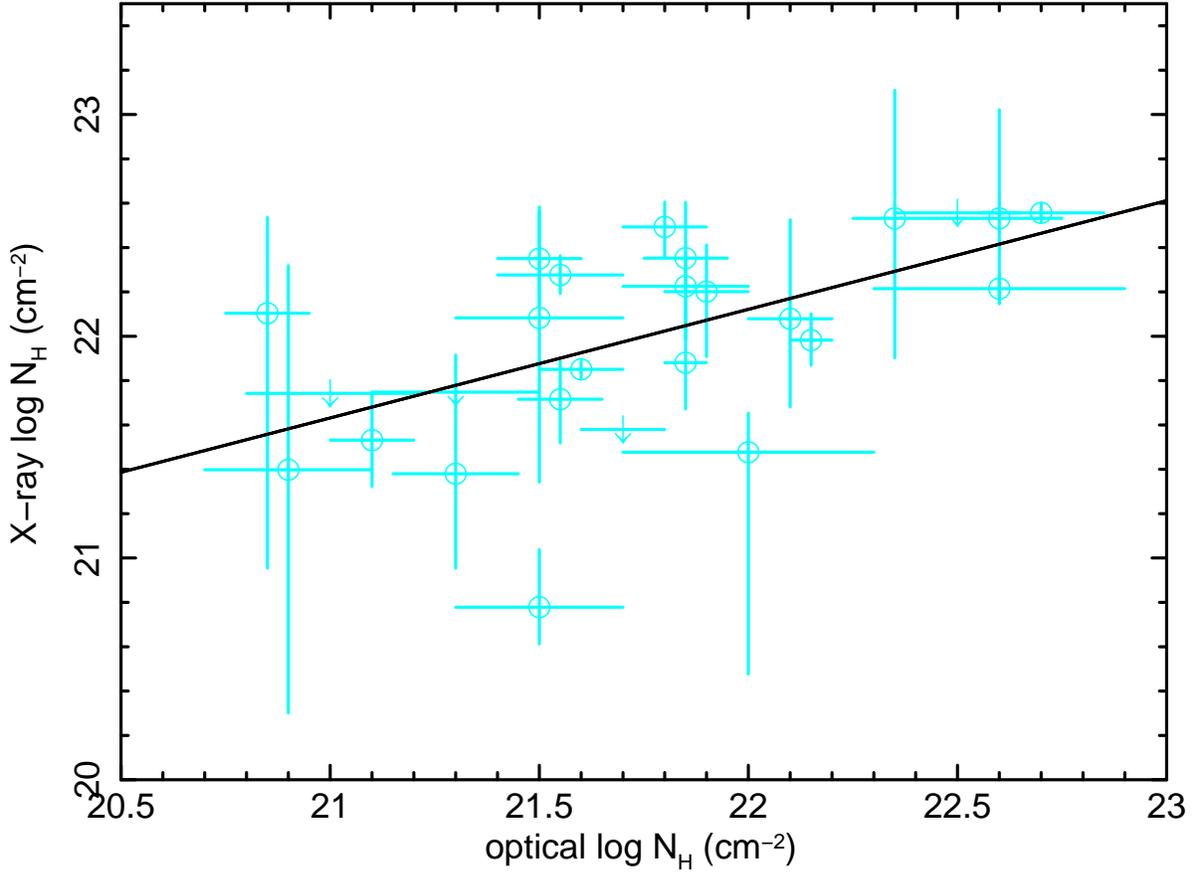}
\caption{Correlation between X-ray absorption $N_{H,x}$ and neutral hydrogen column density $N_{H,opt}$. These bursts (redshifts) are as follows: GRB 050319 (3.240), 
GRB 050401 (2.899), GRB 050730 (3.968), GRB 050820A (2.612), GRB 050922C (2.198), GRB 060115 (3.533), GRB 060206 (4.048), GRB 060210 (3.913), GRB 060707 (3.425), 
GRB 060714 (2.711), GRB 060906 (3.686), GRB 060926 (3.206), GRB 060927 (5.464), GRB 061110B (3.433), GRB 070110 (2.351), GRB 070506 (2.308), GRB 070611 (2.041), 
GRB 070721B (3.628), GRB 070802 (2.455), GRB 071031 (2.692), GRB 080210 (2.641), GRB 080413A (2.433), GRB 080603B (2.690), GRB 080607 (3.037), GRB 080721 (2.591) and 
GRB 080804 (2.205).\label{f6}}
\end{center}
\end{figure}

The quantities of neutral gas in the host galaxies can be obtained from the optical spectra. By the measurement of Ly$\alpha$ absorption, Fynbo et al. (2009) have established
one sample in which 33 values of neutral hydrogen column density $N_{H,opt}$ are derived. The range of these values is from 
$10^{17}cm^{-2}$ to $10^{23}cm^{-2}$(see Fig. 10 of Fynbo et al. 2009), while the true distribution of $N_{H,opt}$ may extend to the higher column densities.
On the other hand, the damped Ly$\alpha$ system with the neutral hydrogen number exceeding $2\times 10^{20}cm^{-2}$ has the possibility of star formation to form a 
protogalaxy (see the simulations by Pontzen et al. 2009 recently); thus, it could be the GRB host. But the thin cloud with the smaller neutral hydrogen values might be  
intervening along the line of sight between the observer and the GRB place; this kind of thin cloud with the column density less than $\sim 10^{20}cm^{-2}$ might not be 
related to the GRB host. Also, in this 
paper, we assume that GRB hosts are rich in neutral gas. Therefore, we only select
the $N_{H,opt}$ values larger than $10^{20}cm^{-2}$ and compare them with the corresponding X-ray absorption $N_{H,x}$ values. We find the relation between X-ray absorption 
and optical neutral gas shown in Fig. \ref{f6}: $log N_{H,x}=(0.49\pm0.04)log N_{H,opt}+(11.3\pm 0.9)$. The linear correlation coefficient is 0.58 with the probability 0.001.
Through this weak relationship, it is likely to find the trace of possible cosmic evolution of neutral gas $N_{H,opt}$, similar to the evolution of X-ray $N_{H,x}$ in Fig. 
\ref{f5}. 
Suppose that GRBs are the unbiased tracers of star formation at high redshift, this possible $N_{H,opt}$ distribution may give an interpretation to the observations of HI 
gas evolution by Prochaska \& Wolfe (2009). However, we caution that the relation between X-ray absorption and optical neutral gas may have larger uncertainties, due to 
the limited redshift range from 2 to 3. A complete sample is required to investigate this relation in the future.

\begin{figure}
\begin{center}
\includegraphics[height=16cm,angle=-90]{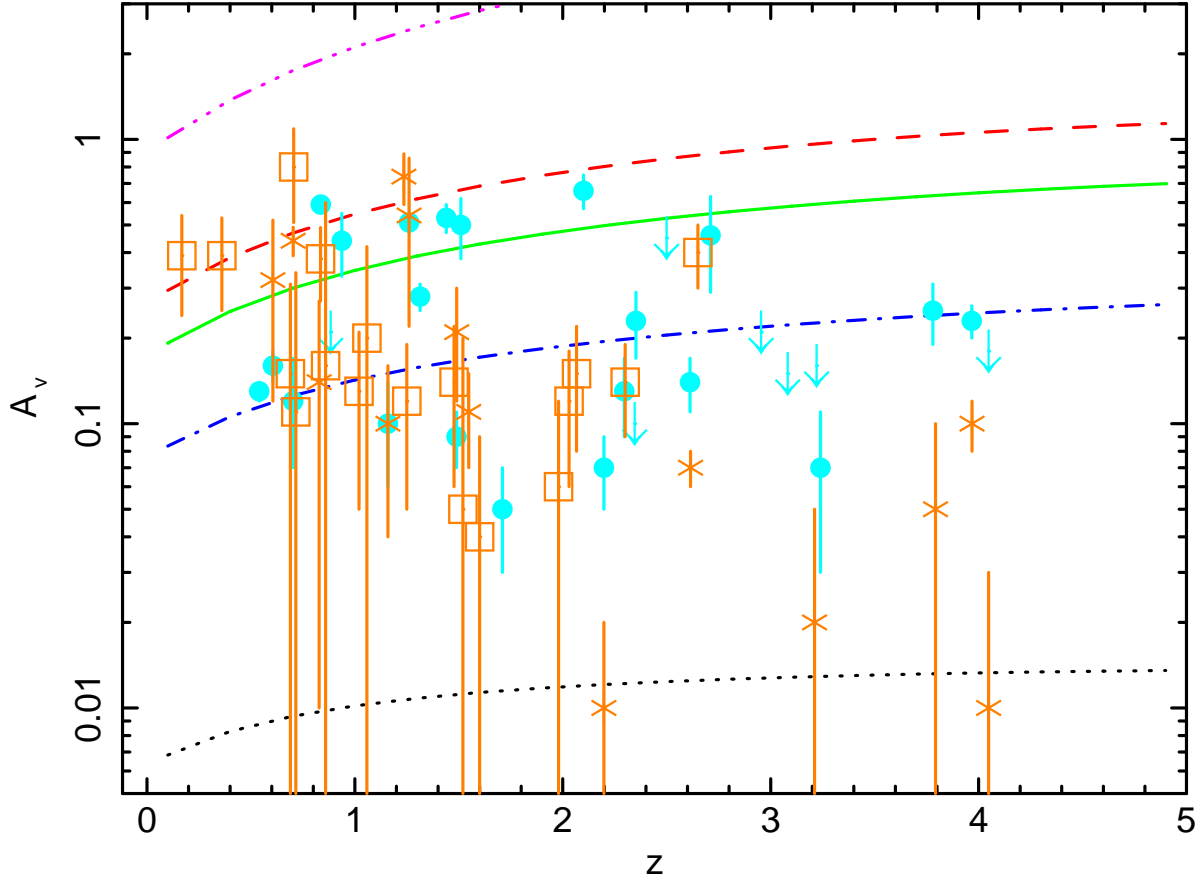}
\caption{Dust absorption $A_v$ distribution with redshift. All the model predictions are the same as those in Fig. 1. The observational data 
taken from Kann et al. (2006) are denoted as squares, from Kann et al. (2007) as '$\ast$', and from Schady et al. (2010) as dots. We note some different $A_v$ values of the 
same burst measured by 
Kann et al. (2007) and Schady et al. (2010): GRB 050730(z=3.967) $A_v1=0.10\pm 0.02$ and $A_v2=0.23\pm 0.02$; GRB 060206(z=4.048) $A_v1=0.01\pm 0.02$ and $A_v2<0.18$; 
GRB 060526(z=3.211) $A_v1=0.02\pm 0.03$ and $A_v2<0.16$; GRB 050820A(z=2.612) $A_v1=0.07\pm 0.01$ and $A_v2=0.14\pm 0.03$; GRB 050922C(z=2.198) $A_v1=0.01\pm 0.01$ and 
$A_v2=0.07\pm 0.02$; GRB 050525A(z=0.606) $A_v1=0.32\pm 0.20$ and $A_v2=0.16\pm 0.02$; GRB 060904B(z=0.703) $A_v1=0.44\pm 0.05$ and $A_v2=0.12\pm 0.05$. $A_v1$ and $A_v2$ 
are the values given by Kann et al. (2007) and Schady et al. (2010) respectively. In general, the values measured by Kann et al. (2007) have larger error bars than those 
measured by Schady et al.(2010). 
 \label{f7}}
\end{center}
\end{figure}

From our model, we see that dust absorption $A_v$ is the function of SFR and metallicity. Combining the effects of both SFR and metallicity, we obtain the redshift 
distribution of dust absorption shown in Fig. \ref{f7}. We see that the Av values from our model are slightly increasing with redshift. From the data of Schady et al. (2010),
we do not find any prominent evidence of $A_v$ variation. But it was claimed by Kann et al. (2007) that the $A_v$ value decreases with increasing redshift. Here, after the 
comparison between the two data sets given by Schady et al. (2010) and Kann et al. (2007), we find that the two data sets are not consistent with each other: some $A_v$ values
of the 
same burst have large difference. The details are listed in the caption to Fig. \ref{f7}. At high redshift, the values of 
Kann et al. (2007) are lower than those of Schady et al. (2010), while at low redshift, the values of Kann et al. (2007) are larger than those of Schady et al. (2010).   
In general, low-mass stars take a long time to evolve to asymptotic giant branch (AGB) 
phase and to produce the dust; thus, AGB population only dominate 
the dust production at local universe. It is suggested that 
at high-redshift the dust factory is supernova explosion. 
However, recently, AGB population has been found to be the 
source of dust production at high redshift 
universe(Valiante et al. 2009): dust production is mainly from
supernova at the beginning of the evolution, but for the time 
larger than $3\times 10^7$ yr, the dust contribution from AGB 
stars increases. 
From the calculation of Valiante et al. (2009), we see that at time 
$1.0\times 10^8$ yr, AGB dust production is still 10 times 
lower than that of supernova. Thus, in our paper, including 
the AGB dust production, the dust extinction $A_v$ is added by 
a factor of 8\%. While at the time $3\times 10^8$ yr, 
the AGB dust production is as same as supernova dust production. 
Thus, we estimate from our model that the total dust extinction 
$A_v$ is 1.7 times of the original value in which only supernova 
production is included. Therefore, we clearly see that both AGB 
and supernova are the origin of dust production at the late 
evolution time larger than $10^8$ yr.
Finally, we cannot ignore the selection effect: at high redshift, the GRBs and their hosts with high absorption values are difficult to be detected by optical telescopes. 

\section{Discussions}
Under the framework of galaxy formation scenario, Lapi et al. (2008) predicted the GRB progenitor rate and redshift distribution. In this paper, without the information of GRB
rates and the cosmological star formation density, we attempt to reveal some properties of GRBs and their host galaxies, which have intrinsic redshift distributions.   
The distributions of these properties with redshift are found to be originally from the star formation in the star-forming galaxies. 
Given a proper galactic evolutionary time and a reasonable dark halo mass, the final results can be obtained by the model calculation. These results are compared with
all kinds of observational data. 

At high redshift, the GRB host galaxy has a plenty of neutral gas, suffering much violent star formation. After the short-time stellar evolution phase, the metal and dust
are released by massive stars; thus, the optical and X-ray GRB emissions will have a strong attenuation locally at high redshift. 
The star formation activity, evolving from relatively massive hosts at high redshift to dwarf galaxies at low redshift, is a coincidence with the so-called downsizing 
scenario(e.g., Heavens et al. 2004). However, at lower redshift, the situation turns to be more complicated. From the morphological statistics by Conselice et al. (2005) and 
Wainwright et al. (2007), the GRB hosts present a broad diversity of galaxy types. About 1/3 host galaxies in the sample of Savaglio et al. (2009) are mergers, while in 
our model the merging and interaction processes are not taken into account. 
In fact, Conselice et al. (2005) found that the GRB hosts at $z>1$ are different from those at $z<1$ in terms of light concentration and the morphological size. Through the 
study of galaxy mass distribution, GRB hosts tracing star formation might be biased at low redshift(Kocevski et al. 2009). It is also complex that the hosts at $z<1$ are not
representative of the general galaxy population(Levesque et al. 2009a). Thus, the properties of these low-redshift GRB hosts presented in this paper could not be reproduced 
by any monolithic process. At least, some low-redshift galaxies may undergo multiple star-forming processes during their whole lifetimes. GRB production can be accompanied 
with any single starburst event.    

From the analysis in this paper, we see that the absorption of GRB X-ray and optical emissions is relatively strong. The strong intrinsic attenuation of GRB host galaxies may 
produce some dark bursts, defined by the index $\beta_{ox}<0.5$, where $\beta_{ox}$ is the flux density ratio between optical and X-ray bands(Jakobsson et al. 2004). 
Rol et al. (2005) proposed several extinction origins from their preliminary results. From our calculations, we see that the heavy attenuation may occur due to 
the following three possibilities: (1) the local environment of the host is metal-enriched, metallicity is higher,  
and/or, the host galaxy in the massive dark halo larger than $10^{12}M_\odot$ may have strong absorption. For example, at redshift 2.5, $Z=1.0Z_\odot$, halo mass 
$M_{halo}=5.0\times 10^{12}M_{\odot}$, after the galactic evolving time $1.0\times 10^8~yr$, we have dust extinction $A_v=1.0$ and the corresponding X-ray absorption 
$N_{H,x}=4.7\times 10^{22}cm^{-2}$; (2) the dust and metals surrounding the GRB in the host galaxy are distributed in an inhomogeneous way; there could be heavy absorption 
through the line of sight, but in other directions the absorption is slight. Also, in our model, we assume that the $A_v$ and $N_{H,x}$ are measured locally and do not change 
significantly if the dust and gas extend out to a few tens to hundreds of pc from the burst(Perna \& Lazzati 2002, D'Elia et al. 2009); however, suppose the observed optical 
extinction is due to the grain absorption far beyond this local region of GRB, the $A_v$-$N_{H,x}$ correlation obtained by Schady et al. (2007, 2010) may be invalid and our 
calculations are strongly biased; (3) as mentioned in Section 2.2.2, the dust produced by the AGB population at high redshift should be taken into account. 

In order to further understand the metal production of GRB environment, we roughly re-estimate the metallicity of GRB hosts under our framework. The mass of metal 
$M_{metal}=SFR\cdot f\cdot f_{dep}\cdot M_{dust}/M_{star}$, where f is the ratio of massive stars to all stars, and $f_{dep}$ is the ratio of mental converted from the dust. 
$f=0.47$ is the case for the stars with the mass larger than $2M_\odot$ by our adopted IMF, $f_{dep}=1.0$ means that all the dust can be transferred to metals. Metallicity is 
defined by $Z=M_{metal}/M_{gas}$. From the SFR calculated by Granato et al. (2004) and Mao et al. (2007), as an example, at redshift 6, we obtain the metallicity as 
$Z\sim 2.75\times 10^{-2}(M_{dust}/M_{star})$. If we take a supernova with the dust production of $10^{-3}$ solar mass(Pozzo et al. 2004), we obtain the upper limit of 
metallicity $Z\sim 10^{-3}Z_{\odot}$, which is lower than the measurement($Z>0.02Z_\odot$) of GRB 050904(Campana et al. 2007). If we take the dust mass 
0.08-0.3 solar mass per primordial massive supernova(Todini \& Ferrara 2001), we have the result which is consistent with the observation. The estimation 
values of Population I/II metallicity are lower than the observational values at high redshift, meaning that the imprints from some primordial objects(Kawai et al. 2006), such
as pop III stars and mini-quasars, have to be included in the possible cosmic evolution properties of these GRB host galaxies.
According to this estimation, the metal-enriched environment of GRB host galaxy naturally gives the reason of the strong attenuation in X-ray and optical band measurements. 
In our model, the initial galactic evolutionary time of about $10^7$ yr of host galaxies is given, but the corresponding metallicity about 
$Z\sim 0.3Z_\odot$(Lapi et al. 2008) is not a necessary condition, as mentioned by Levesque et al. (2009b) that low metallicity may not be required for a relativistic 
explosion. 
With our model, the massive dark halo above $10^{12}M_\odot$ can host the GRB galaxy in which the metallicity is relatively high, although most GRB host galaxies are inside 
the dark halos with the masses less than $10^{12}M_\odot$. On the other hand, a host galaxy with a top-heavy IMF, meaning that much more massive stars are 
involved, can produce more metals in relatively short time during the galactic evolution phase. For instance, the Wolf-Rayet star with a mass of 80$M_\odot$ and
initial metallicity $Z=0.001$ has the possibility to self-enrich the HII region(Kr\"{o}ger, Hensler \& Freyer 2006) and to produce the GRB event(Eldridge et al. 2006).

In this paper, we have calculated the SFR, galactic mass, and metallicity of the GRB host galaxies. The absorption variations with redshift in the X-ray and optical bands are
presented as well. Some selection effects have been taken into account through our calculation. Other observational biases should also be considered. All the redshift 
measurements 
come from the optical observations so that some optical-faint GRBs and host galaxies are ignored. Moreover, at high redshift only most luminous galaxies with high SFRs 
can be detected, indicating that some low-luminosity cases are not included. However, our calculations come from the intrinsic star formation of GRB host galaxies. 
Thus, due to all these selection effects and observational bias mentioned in the paper, the intrinsic properties of GRB afterglows and the hosts
by the model calculations have some differences to those from observations. 
As SFR evolution plays a dominant role in the calculations, compared to the situation at low redshift, in general, star formation in the metal-poor environment 
at high redshift may provide more powerful GRB explosion. Therefore, although the effective threshold is given by Kistler et al. (2009), we speculate 
that improving the sensitivity of detectors on the high-energy telescopes is not strongly useful to catch more high-redshift but faint GRBs, since low-energy-released GRBs 
are almost absent in the high-redshift universe.          

\acknowledgments
We are grateful to Dr. Salvaterra, R. and Campana S. for their helpful discussion. We thank the referee to give us the constructive suggestions. This work is supported by the 
following research grants (P.I. Guido Chincarini): ASI grant Swift I/011/07/0, by the Ministry of University and Research of Italy (PRIN MIUR 2007TNYZXL), by MAE and by the 
University of Milano Bicocca (Italy). 

%{\it Facilities:} \facility{Swift}

%\clearpage

\clearpage

\clearpage

\clearpage

\clearpage

\clearpage

%\begin{figure}
%\begin{center}
%\includegraphics[height=16cm,angle=-90]{rz.ps}
%\caption{The evolution of the ratio between X-ray absorption $N_{H,x}$ and optical $A_v$ from the {\it Swift} observational data(Schady et al. 2009). $A_v$ values are 
%selected by the measurement of Small Magellanic Clouds(SMC) dust-extinction law, as the results of both power-law and broken power-law continuum fitting are given for each 
%burst, we take the $A_v$ value with better null hypothesis probability. Solid line is the optimized solution by the linear regression.\label{f5}}
%\end{center}
%\end{figure}

\clearpage

\clearpage

\clearpage

\clearpage


\begin{thebibliography}{}
\bibitem[Akerlof \& Swan(2007)]{Akerlof07} Akerlof, C. W. \& Swan, H. F. 2007, \apj, 671, 1868
\bibitem[Campana et al.(2007)]{Campana07} Campana, S., et al. 2007, \apj, 654, L17
\bibitem[Campana et al.(2010)]{Campana10} Campana, S., Th\"{o}ne, C. C., de Ugarte Postigo, A., Tagliaferri, G., Moretti, A., \& Covino, S. 2010, \mnras, 402, 2429
\bibitem[Campisi et al.(2009)]{Campisi09} Campisi, M. A., De Lucia, G., Li, L.-X., Mao, S., \& Kang, X. 2009, arXiv:astro-ph/ 0908.2427
\bibitem[Calzetti et al.(2000)]{Calzetti00} Calzetti, D., Armus, L., Bohlin, R. C., Kinney, A. L., Koornneef, J., \& Storchi-Bergmann, T. 2000, \apj, 533, 682
\bibitem[Castro Cer\'{o}n et al.(2008)]{CC08} Castro Cer\'{o}n, J. M., Michalowski, M. J., Hjorth, J., Malesani, D., Gorosabel, J., Watson, D., \& Fynbo, J. P. U. 2008, 
arXiv:astro-ph/0803.2235
\bibitem[Chen, Li \& Wei(2006)]{Chen06} Chen, S. L., Li, A. \& Wei, D .M. 2006, \apj, 647, L13
\bibitem[Christensen, Hjorth \& Gorosabel(2004)]{Christensen04} Christensen, L., Hjorth, J. \& Gorosabel, J. 2004, \aap, 425, 913
\bibitem[Cirasuolo et al.(2005)]{Cirasuolo05} Cirasuolo, M., Shankar, F., Granato, G. L., De Zotti, G., \& Danese, L. 2005, \apj, 629, 816 
\bibitem[Conselice et al.(2005)]{Conselice05} Conselice, C. J., et al. 2005, \apj, 633, 29
\bibitem[Courty, Bjornsson \& Gudmundsson(2007)]{Courty07} Courty, S., Bjornsson, G. \& Gudmundsson, E. H. 2007, \mnras, 376, 1375
\bibitem[D'Elia et al.(2009)]{D'Elia09} D'Elia, V., et al. 2009, \aap, 503, 437
\bibitem[Eldridge et al.(2006)]{Eldridge06} Eldridge, J. J., Genet, F., Daigne, F., \& Mochkovitch, R. 2006, \mnras, 367, 186
\bibitem[Evans et al.(2009)]{Evans09} Evans, P., et al. 2009, \mnras, 397, 1177
\bibitem[Fynbo et al.(2006)]{Fynbo06} Fynbo, J. P. U., et al. 2006, \aap, 451, L47 
\bibitem[Fynbo et al.(2009)]{Fynbo09} Fynbo, J. P. U., et al. 2009, arXiv:astro-ph/0907.3449
\bibitem[Granato et al.(2004)]{Granato04} Granato, G. L., De Zotti, G., Silva, L., Bressan, A., \& Danese, L. \apj, 2004, 600, 580 
\bibitem[Granato et al.(2006)]{Granato06} Granato, G. L., Silva, L., Lapi, A., Shankar, F., De Zotti, G., \& Danese, L. \mnras, 2006, 368, L72
\bibitem[Han et al.(9010)]{Han10} Han, X. H., Hammer, F., Liang, Y. C., Flores, H., Rodrigues, M., Hou, J. L. \& Wei, J. Y. 2010, arXiv:astro-ph/1001.2476
\bibitem[Heavens et al. 2004)]{heavens04} Heavens, A., Panter, B., Jimenez, R. \& Dunlop, J. 2004, Nature, 428, 625 
\bibitem[Hjorth et al.(2003)]{Hjorth03} Hjorth, J., et al. 2003, Nature, 423, 847
\bibitem[Hopkins \& Beacom(2006)]{Hopkins06} Hopkins, A. M. \& Beacom, J. F. 2006, \apj, 651, 142
\bibitem[Jakobsson et al.(2004)]{Jakobsson04} Jakobsson, P., Hjorth, J., Fynbo, J. P. U., Watson, D., Pedersen, K., Bj\"{o}rnsson, G., \& Gorosabel, J. 2004, \apj, 617, L21
\bibitem[Jakobsson et al.(2005)]{Jakobsson05} Jakobsson, P., et al. 2005, \mnras, 362, 245 
\bibitem[Kann et al.(2006)]{Kann06} Kann, A., Klose, S. \& Zeh, A. 2006, \apj, 641, 993
\bibitem[Kann et al.(2007)]{Kann07} Kann, A., et al. 2007, arXiv:astro-ph/0712.2186
\bibitem[Kawai et al.(2006)]{Kawai06} Kawai, N., et al. 2006, Nature, 440, 184
\bibitem[Kewley et al.(2007)]{Kewley07} Kewley, L. J., Brown, W. R., Geller, M. J., Kenyon, S. J., \& Kurtz, M. J. 2007, \aj, 133, 882
\bibitem[Kistler et al.(2009)]{Kistler09} Kistler, M. D., Y\"{u}ksel, H., Beacom, J. F., Hopkins, A. M., \& Wyithe, J. S. B. 2009, arXiv:astro-ph/0906.0590
\bibitem[Kocevski, West \& Modjaz(2009)]{Kocevski09}Kocevski, D., West, A. A. \& Modjaz, M. 2009, arXiv:astro-ph/0905.1953
\bibitem[Kr\"{o}ger, Hensler,\& Freyer(2006)]{Kroger06} Kr\"{o}ger, D., Hensler, G. \& Freyer, T. 2006, \aap, 450, L5
\bibitem[Kumar, Narayan \& Johnson(2008)]{Kumar08} Kumar, P., Narayan, R. \& Johnson, J. L. 2008, Science, 321, 376 
\bibitem[Lapi et al.(2006)]{Lapi06} Lapi, A., Shankar, F., Mao, J., Granato, G. L., Silva, L., De Zotti, G., \& Danese, L. 2006, \apj, 650, 42
\bibitem[Lapi et al.(2008)]{Lapi08} Lapi, A., Kawakatu, N., Bosnjak, Z., Celotti, A., Bressan, A., Granato, G. L., \& Danese, L. 2008, \mnras, 386, 608
\bibitem[Levesque et al.(2009a)]{Levesque09a}Levesque, E. M., Berger, E., Kewley, L. J., \& Bagley, M, M. 2009a, arXiv:astro-ph/0907.4988
\bibitem[Levesque et al.(2009b)]{Levesque09b}Levesque, E. M., et al. 2009b, arXiv:astro-ph/0908.2818
\bibitem[Li(2008)]{Li08} Li, L.-X. 2008, \mnras, 388, 1487
\bibitem[Li, Li \& Wei(2008)]{LiLiWei08} Li, Y., Li, A. \& Wei, D. M. 2008a, \apj, 678, 1136
\bibitem[Li et al.(2008)]{Lietal08} Li, A., Liang, S. L., Kann, D. A., Wei, D. M., Klose, S. \& Wang, J. Y. 2008b, \apj, 685, L1046
\bibitem[Malesani et al.(2004)]{Malesani04} Malesani, D., et al. 2004, \apj, 609, L5
\bibitem[Malesani et al.(2009)]{Malesani09} Malesani, D., Hjorth, J., Fynbo, J. P. U., Milvang-Jensen, B., Jakobsson, P., \& Jaunsen, A. O. 2009, AIPC, 1111, 513
\bibitem[Mao et al.(2007)]{Mao07} Mao, J., Lapi, A., Granato, G. L., de Zotti, G., \& Danese, L. 2007, \apj, 667, 655 
\bibitem[Mazzali et al.(2006)]{Mazzali06} Mazzali, P. A., et al. 2006, Nature, 442, 1018
\bibitem[Metzger et al.(1997)]{Metzger97} Metzger, M. R., Djorgovski, S. G., Kulkarni, S. R., Steidel, C. C., Adelberger, K. L., Frail, D. A., Costa, E., \& Frontera, F. 1997,
Nature, 387, 878
\bibitem[Nardini et al.(2009)]{Nardini09} Nardini, M., Ghisellini, G., Ghirlanda, G., \& Celotti, A. 2009, arXiv:astro-ph/0907.4157
\bibitem[Paczynski(1998)]{Paczynski98} Paczynski, B. 1998, ApJ, 494, L45
\bibitem[Perley et al.(2009)]{Perley09} Perley, D. A., et al. 2009, \aj, 138, 1690
\bibitem[Perna \& Lazzati(2002)]{Perna02} Perna, R. \& Lazzati, D. 2002, \apj, 580, 261
\bibitem[Pontzen et al.(2009)]{Pontzen09} Pontzen, A., et al. 2009, arXiv:astro-ph/0909.1321
\bibitem[Pozzo et al.(2004)]{Pozzo04} Pozzo, M., Meikle, W. P. S., Fassia, A., Geballe, T., Lundqvist, P., Chugai, N. N. \& Sollerman, J. 2004, \mnras, 352, 457
\bibitem[Prochaska \& Wolfe 2009]{Prochaska09} Prochaska, J. X. \& Wolfe, A. M. 2009, \apj, 696, 1543
\bibitem[Rol et al.(2005)]{Rol05} Rol, E., Wijers, R. A. M. J., Kouveliotou, C., Kaper, L., \& Kaneko, Y. 2005, \apj, 624, 868
\bibitem[Romano et al.(2002)]{Romano02} Romano, D., Silva, L., Matteucci, F., \& Danese, L. 2002, \mnras, 334, 444
\bibitem[Salvaterra et al.(2009a)]{Salvaterra09a} Salvaterra, R., et al. 2009a, Nature, 461, 1258
\bibitem[Salvaterra et al.(2009b)]{Salvaterra09b} Salvaterra, R., Guidorzi, C., Campana, S., Chincarini, G., \& Tagliaferri, G. 2009b, \mnras, 396, 299
\bibitem[Savaglio et al.(2005)]{Savaglio05} Savaglio, S., et al. 2005, \apj, 635, 260
\bibitem[Savaglio(2006)]{Savaglio06} Savaglio, S. 2006, New J. Phys., 8, 195
\bibitem[Savaglio, Glazebrook \& Le Borgne(2009)]{Savaglio09} Savaglio, S., Glazebrook, K. \& Le Borgne, D. 2009, \apj, 691, 182
\bibitem[Schady et al.(2007)]{Schady07} Schady, P., et al. 2007, \mnras, 377, 273
\bibitem[Schady et al.(2010)]{Schady09} Schady, P., et al. 2010, \mnras, 401, 2773
\bibitem[Stanek et al.(2003)]{Stanek03} Stanek, K. Z., et al. 2003, \apj, 591, L17
\bibitem[Tanvir et al.(2009)]{Tanvir09} Tanvir, N. R., et al. 2009, Nature, 461, 1254
\bibitem[Th\"{o}ne et al.(2008)]{Thone08} Th\"{o}ne, C. C., et al. 2008, \apj, 676, 1151
\bibitem[Todini \& Ferrara (2001)]{Todini01} Todini, P. \& Ferrrara, A. 2001, \mnras, 325, 726
\bibitem[Valiante et al.(2009)]{valiante09} Valiante, R., Schneider, R., Bianchi, S. \& Andersen, A. C. 2009, \mnras, 397, 1661
\bibitem[van Paradijs et al.(1997)]{van97} van Paradijs, J., et al. 1997, Nature 386, 686 
\bibitem[Wainwright, Berger \& Penprase(2007)]{Wainwright07} Wainwright, C., Berger, E. \& Penprase, B. E. 2007, \apj, 657, 367
\bibitem[Wiersema et al.(2007)]{Wiersema07} Wiersema, K., et al. 2007, \aap, 464, 529
\bibitem[Woosley(1993)]{Woosley93} Woosley, S. E. 1993, \apj, 405, 273
\bibitem[Xu, Zou \& Fan(2008)]{Xu08} Xu, D., Zou, Y.-C. \& Fan, Y.-Z. 2008, arXiv:astro-ph/0801.4325
\bibitem[Yan et al.(2009)]{Yan09} Yan, H., Windhorst, R., Hathi, N., Cohen, S., Ryan, R., O'Connell, R., \& McCarthy, P. 2009, arXiv:astro-ph/0910.0077
\bibitem[Y\"{u}ksel et al. 2008]{Yuksel08} Y\"{u}ksel, H., Kistler, M. D., Beacom, J. F., \& Hopkins, A. M. 2008, \apj, 683, L5
\bibitem[Zheng, Deng \& Wang (2009)]{Zheng09} Zheng, W., Deng, J. \& Wang, J. 2009, arXiv:astro-ph/0906.2244
\end{thebibliography}
\end{document}